\documentclass[11pt,a4paper]{article}
\pdfoutput=1
\usepackage{jheppub}
\usepackage{amsfonts}
\usepackage{bbm}
\usepackage{graphicx}
\usepackage{caption}
\usepackage{subcaption}
\usepackage{bigints}
\usepackage[normalem]{ulem}
\usepackage{slashed}

 \def\g{\gamma}

\def\e{\epsilon}

\def\m{\mu}
\def\n{\nu}

\newcommand{\RN}[1]{%
  \textup{\uppercase\expandafter{\romannumeral#1}}%
}

\newcommand{\nn}{\nonumber}

\def\e{\epsilon}

\def\p{\partial}

\def\bea{\begin{eqnarray}}
\def\eea{\end{eqnarray}}
\def\be{\begin{equation}}
\def\ee{\end{equation}}
\def\ba{\begin{align}}
\def\ea{\end{align}}

\newcommand{\bem}{\begin{pmatrix}}
\newcommand{\eem}{\end{pmatrix}}

\def\={\;  = \;}
\def\+{\, + \,}

\def\rt2{\sqrt{2}}

\title{Non-relativistic Conformal Field Theory in the Presence of Boundary}

\author{\small{Rajesh Kumar Gupta,}}
\author{\small{Ramanpreet Singh}}
\affiliation{\small{Department of Physics, Indian Institute of Technology Ropar,
Rupnagar, Punjab 140001, India}}

\emailAdd{rajesh.gupta@iitrpr.ac.in, singhramanpreet338@gmail.com}

\abstract{ We study non-relativistic conformal field theory on a flat space in the presence of a planar boundary. We compute correlation functions of primary operators and obtain the expression for the boundary conformal block. We also discuss the non-relativistic conformal field theory on a general curved background in the presence of a boundary. As an example, we discuss the spectrum of boundary primary operator and compute scaling dimensions in a fermionic theory near one and three spatial dimensions.
}

\makeatletter
\gdef\@fpheader{}
\makeatother

\begin{document}

%

\maketitle

\section{Introduction \label{sec:intro}}
A relativistic conformal field theory has a special significance in the space of quantum field theories. It appears as the description of the fixed point of the renormalization group flow in a relativistic quantum field theory.  The symmetry group of a relativistic conformal field theory is much bigger than the Poincare invariance of the relativistic quantum field theory. 
It is the maximal invariance group of a free massless Klein-Gordon equation. It consists of the Poincare transformations and dilatation and special conformal transformations.

Following the symmetry argument, one can defines a non-relativistic analogue of conformal field theories. A non-relativistic conformal field theory is a field theory with the Schr$\ddot{\text{o}}$dinger group as the symmetry group. 
The Schr$\ddot{\text{o}}$dinger group is the maximal invariance group of the free Schr$\ddot{\text{o}}$dinger equation. It consists of the centrally extended Galilean group together with dilatation and an expansion transformation~\cite{Hagen:1972pd, Niederer:1972}. Note that a Schr$\ddot{\text{o}}$dinger group in $(d+1)$-dimensional space-time is a smaller group than the $(d+1)$-dimensional relativistic conformal group. In three spatial dimensions, the former is a twelve parameters group (plus the central extension), whereas the latter has fifteen parameters.
One of the motivations to study such theories is that there are ultra cold systems that can be designed experimentally whose properties can be described by a non-relativistic conformal field theory. One such case is the fermi gas at unitarity, where the system can be fine-tuned to exhibit the scale invariance.

Non-relativistic conformal field theories with Schr$\ddot{\text{o}}$dinger symmetry group have been studied previously~\cite{Henkel:1993sg, Mehen:1999nd, Henkel:2003pu, Nishida:2006br, Nishida:2006eu, Nishida:2007pj, Braaten:2008uh, Golkar:2014mwa, Goldberger:2014hca}. Schr$\ddot{\text{o}}$dinger invariance fixes one and two point functions whereas three and higher point functions are fixed up to a function of Schr$\ddot{\text{o}}$dinger invariant cross ratios. As in a relativistic CFTs, Schr$\ddot{\text{o}}$dinger invariant theory also exhibits state operator correspondence where the scaling dimension of a primary operator maps to energy eigen value of the non-relativistic system in a harmonic potential.
A convergent operator product expansion also exists, which expresses the product of two primary operators with given particle numbers as a sum of conformal multiplets consistent with the particle number conservation. This paper is interested in studying some of these properties for a non-relativistic conformal field theory on a manifold with a boundary.

Boundaries have played an important role in several areas of theoretical physics ranging from condensed matter physics to string theory. In condensed matter physics, systems exhibit physically interesting phenomena due to the presence of a boundary.
An example of this is the topological insulator, where there is symmetry protected conducting Dirac fermions at the surface of the otherwise insulator. For a statistical system with a boundary, new critical exponents appear near the critical points, which are unrelated to the bulk critical exponents. In the holographic duality, a weakly interacting theory of gravity describes the strongly interacting quantum fields living on the conformal boundary of anti-de Sitter space. D-branes that act as a boundary for an open string to end provide a non-perturbative window into string theory.

The presence of the boundary breaks the spacetime symmetry of the underlying system. In the case of conformal symmetry, the boundary conditions break the conformal group to a subgroup of transformations that preserve the boundary conditions.
Relativistic conformal field theories on a manifold with a boundary have been studied extensively~\cite{Affleck:1991tk, Cardy:1991tv, McAvity:1993ue, McAvity:1995zd, Friedan:2003yc, Cardy:2004hm, Fursaev:2015wpa, Solodukhin:2015eca, Herzog:2015ioa, Jensen:2015swa, Herzog:2017xha, Casini:2018nym}. These explorations have led to deeper insights into systems with a boundary, e.g. the critical behaviour, the renormalization group flow, boundary monotonicity theorems, and anomalies. Little work has been done in this direction in the context of non-relativistic field theories.
In the present article, we initiate the study of non-relativistic conformal field theories on a manifold with a boundary. 
The presence of a boundary breaks the Schr$\ddot{\text{o}}$dinger group to the non-relativistic transformations which preserve the boundary. In the case of a non-relativistic conformal field theory on a flat space with a planar boundary, the reduced group includes the centrally extended Galilei group of the boundary, dilatations and expansion transformation. We study the consequences of the reduced symmetry group on the correlation functions and the boundary operator product expansion.

The organization of this paper is as follows. In  section~\ref{SymmetriesandBoundary}, we review the Schr$\ddot{\text{o}}$dinger group and algebra in $d$-spatial dimensions. We then discuss the effect of planar boundary and the resultant symmetry algebra. In section~\ref{Correlationfunctions}, we compute the one and the two-point function of primary operators in the presence of the planar boundary. For this, we solve the Ward identities of the Schr$\ddot{\text{o}}$dinger group satisfied by the correlation function. We also obtain the boundary operator product expansion of the bulk primary operators. In section~\ref{NRCFTonCurvedSpace}, we discuss the coupling of the non-relativistic conformal field theory to non-trivial background fields very briefly. We write down the diffeomorphic and Weyl invariant action of a Schr$\ddot{\text{o}}$dinger invariant Lagrangian in the presence of a boundary. In section~\ref{Freefermions}, we construct an explicit example of an interacting non-relativistic conformal field theory in the presence of a boundary. The non-relativistic conformal field theory describes the fixed point of the free fermions with interactions localized at the boundary. 
\section{Schr$\ddot{\text{o}}$dinger symmetries and boundary}\label{SymmetriesandBoundary}
The relativistic conformal group in $(d+1)$-dimensional spacetime, $SO(d+1,2)$, is a group of spacetime transformations which keep the massless Klein-Gordon equation invariant. 
The Schr$\ddot{\text{o}}$dinger group is the group of space-time transformations which leaves the free Schr$\ddot{\text{o}}$dinger equation invariant. Let $\hat S(t,\vec x)$ is the Schr$\ddot{\text{o}}$dinger operator acting on a wave function $\psi(t,\vec x)$,
\be\label{SchrodingerEq}
\hat S(t,\vec x)\psi(t,\vec x)=0\,.
\ee
Then, the space-time coordinate transformation $(t,\vec x)\rightarrow g(t,\vec x)$ is said to be the symmetry transformation of the Schr$\ddot{\text{o}}$dinger equation if there exists a (non-trivial) linear transformation on the wave function~\cite{Niederer:1972}, i.e.
\be
\psi(t,\vec x)\rightarrow (T_{g}\psi)(t,\vec x)=f_{g}(g^{-1}(t,\vec x))\psi(g^{-1}(t,\vec x))\,,
\ee
with the property that transformed wave function, i.e. $(T_{g}\psi)(t,\vec x)$, is also a solution of the Schr$\ddot{\text{o}}$dinger equation~\eqref{SchrodingerEq}. The solution $(g,f_{g})$ gives rise to the Schr$\ddot{\text{o}}$dinger group. The Schr$\ddot{\text{o}}$dinger group contains Galilean group and the group of dilatation and expansion transformation.
The Lie algebra of the Schr$\ddot{\text{o}}$dinger group $\mathfrak{sch}(d)$ in $d$-spatial dimensions consists of centrally extended Galilean algebra together with the algebra involving the generators of dilatation and expansion transformations. 
The non-zero commutation relations of the Hermitian generators of the Schr$\ddot{\text{o}}$dinger algebra are,
\bea
&&[M_{IJ},M_{KL}]=i(\delta_{IK}M_{JL}-\delta_{JK}M_{IL}-\delta_{IL}M_{JK}+\delta_{JL}M_{IK})\,,\nn\\
&&[M_{IJ},K_{K}]=i(\delta_{IK}K_{J}-\delta_{JK}K_{I}),\quad [M_{IJ},P_{K}]=i(\delta_{IK}P_{J}-\delta_{JK}P_{I})\,,\nn\\
&&[P_{I},K_{J}]=-i\delta_{IJ}M,\quad [H,K_{I}]=-iP_{I}\,,\nn\\
&&[D,P_{I}]=iP_{I},\quad [P_{I},C]=-iK_{I},\quad [D,K_{I}]=-iK_{I}\,,\nn\\
&&[D,C]=-2iC,\quad [D,H]=2iH,\quad [C,H]=-iD\,.
\eea
In the above, $M_{IJ}$, $P_{I}$, $K_{I}$ and $H$, for $I,J=1,..,d$, are the generators of rotation, spatial translation, boost and time translation, respectively and constitute the group of Galilean transformations. The central extension of the Galilean group by $U(1)$ is generated by the particle number generator $M$. The rest of the generators for dilations and expansion transformation are denoted by $D$ and $C$, respectively.
The action of these on the space-time coordinates are given by 
\bea
&&\text{Galilean transformation:} \,\,\,\,\,\quad\,x_{I}\rightarrow R_{IJ} x_{J}+v_{I}\,t+a_{I},\quad t\rightarrow t+b\,,\nn\\
&&\text{Dilatation transformation:} \,\quad(x_{I},t)\rightarrow (\lambda\, x_{I},\lambda^{2}t)\,,\nn\\
&&\text{Expansion transformation:} \,\quad(x_{I},t)\rightarrow \Big(\frac{ x_{I}}{1+\beta\, t},\frac{t}{1+\beta\, t}\Big)\,.
\eea
A non-relativistic field theory invariant under the action of the Schr$\ddot{\text{o}}$dinger group is the non-relativistic analog of conformal field theory. We will call such theory non-relativistic conformal field theory.

In this paper, we will be interested in a non-relativistic conformal field theory in the presence of boundary. Clearly, the space-time transformation under the action of the Schr$\ddot{\text{o}}$dinger group does not preserve the boundary. For example, the presence of boundary breaks the translation and boost symmetry transformations perpendicular to the boundary. Therefore, in the presence of boundary not all the symmetry transformations are maintained. Let us denote the $(d+1)$-dimensional space-time coordinates by $(t,x)$ where $x=(\vec {\bf x},y)$ and $\vec{\bf x}$ is the boundary coordinate. The boundary is at $y=0$. The symmetry group that we will consider is the subgroup of the Schr$\ddot{\text{o}}$dinger group whose action on the space-time coordinates leaves $y=0$ invariant. These will be generated by $(d-1)$-dimensional translation and rotation, $(d-1)$-dimensional boost, time translation, scale and expansion transformations. The action of these on the space-time coordinates are given by (except the particle number generator)
\be
\vec {\bf x}\rightarrow R\,\vec {\bf x}+\vec v\,t+\vec a,\quad t\rightarrow t+b\,.
\ee
The action of the scale transformation is
\be
D: ((\vec{\bf x},y),t)\rightarrow ((\lambda\,\vec{\bf x},\lambda\, y),\lambda^{2}t)\,.
\ee
Finally, the action of the expansion generator is
\be
C: ((\vec{\bf x},y),t)\rightarrow \Big(\Big(\frac{\vec{\bf x}}{1+\beta t},\frac{y}{1+\beta t}\Big),\frac{t}{1+\beta t}\Big)\,.
\ee
The generators of the reduced Schr$\ddot{\text{o}}$dinger group in the presence of the boundary satisfy the following algebra
\bea
&&[M_{ij},M_{kl}]=i(\delta_{ik}M_{jl}-\delta_{jk}M_{il}-\delta_{il}M_{jk}+\delta_{jl}M_{ik})\,,\nn\\
&&[M_{ij},K_{k}]=i(\delta_{ik}K_{j}-\delta_{jk}K_{i}),\quad [M_{ij},P_{k}]=i(\delta_{ik}P_{j}-\delta_{jk}P_{i})\,,\nn\\
&&[P_{i},K_{j}]=-i\delta_{ij}M,\quad [H,K_{i}]=-iP_{i}\,,\nn\\
&&[D,P_{i}]=iP_{i},\quad [P_{i},C]=-iK_{i},\quad [D,K_{i}]=-iK_{i}\,,\nn\\
&&[D,C]=-2iC,\quad [D,H]=2iH,\quad [C,H]=-iD\,.
\eea
In the above $i,j=1,..,(d-1)$ are the boundary indices. Note that in the case of $d=1$ i.e. Schr$\ddot{\text{o}}$dinger invariant theory on one spatial dimension in the presence of boundary, there are no boost and spatial translation symmetry. As a result, in this case, we do not have the $U(1)$ symmetry generated by the generator $M$. The relevant symmetry group for $d=1$ is $SL(2,\mathbb R)$ which is generated by $D,C$ and $H$.
\section{Correlation function and boundary OPE}\label{Correlationfunctions}
In this section, we compute the one and two-point correlation function of a scalar primary operator using the Ward identities. We find that the symmetry group in the presence of the boundary is powerful enough to fix the one point function whereas the two-point function is fixed up to an undetermined function of a cross ratio. 
\subsection{Bulk one and two-point function}
The action of the Schr$\ddot{\text{o}}$dinger group on a bulk scalar primary operator $\mathcal O(t,x)$ is given by
\bea
&&[M,\mathcal O(t,x)]=m_{\mathcal O}\mathcal O(t,x),\qquad [D,\mathcal O(t,x)]=i(2t\p_{t}+x_{I}\p_{I}+\Delta_{\mathcal O})\mathcal O(t,x)\,,\nn\\
&&[C,\mathcal O(t,x)]=(-it^{2}\p_{t}-itx_{I}\p_{I}-it\Delta_{\mathcal O}+\frac{m_{\mathcal O}}{2}x^{2})\mathcal O(t,x)\,,\nn\\
&&[P_{i},\mathcal O(t,x)]=i\p_{i}\mathcal O(t,x)\,,\quad [H,\mathcal O(t,x)]=-i\p_{t}\mathcal O(t,x)\,,\nn\\
&&[K_{i},\mathcal O(t,x)]=(-it\p_{i}+m_{\mathcal O}x_{i})\mathcal O(t,x)\,.
\eea
In the above $I,J=1,...,d$ are the bulk indices and $i,j=1,...,(d-1)$ are the boundary indices. We will use the above commutation relations to fix the one and two-point function of the scalar primary operator. Note that $d=1$ requires a special treatment since in this case, we do not have the centrally extended Galilean symmetry.

Let's start with the one point function. We note that the Ward identity for the $U(1)$ generator $M$ forces the one point function of any scalar operator to zero unless $m_{\mathcal O}=0$. This is because
\be
<[M,\mathcal O(t,x)]>=0,\quad\Rightarrow \quad m_{\mathcal O}=0\,.
\ee
Thus the one point function of a scalar primary operator, consistent with translation, rotation and scale invariance is
\be
<\mathcal O(t,x)>=\frac{c\,\delta_{m_{\mathcal O},0}}{y^{\Delta_{\mathcal O}}}\,,
\ee
where $c$ is a constant.

Next, we consider the two point function of scalar primary operators. The Ward identity for the $U(1)$ generator is
\be\label{MassCondition}
<\mathcal T[M,\mathcal O_{1}(t_{1},x_{1})]\mathcal O_{2}(t_{2},x_{2})>+<\mathcal T\mathcal O_{1}(t_{1},x_{1})[M,\mathcal O_{2}(t_{2},x_{2})]>=0\,.
\ee
In the above $\mathcal T$ is the notation for the time ordered product. The above relation is satisfied provided $m_{\mathcal O_{1}}=-m_{\mathcal O_{2}}$. 

Invariance under space and time translations, and spatial rotations imply that the correlation function depends on $|\vec x_{12}|=|\vec x_{1}-\vec x_{2}|$ and $t_{12}=(t_{1}-t_{2})$. 
Furthermore, we note that $t_{12}$ is invariant under all symmetries except expansion and scale transformation. Under the expansion transformation, it transforms as
\be
t_{12}\rightarrow \frac{t_{12}}{(1+\beta t_{1})(1+\beta t_{2})}\,,
\ee
where $\beta$ is the transformation parameter. We see that one can construct a cross ratio which is invariant under all the symmetries and is given by
\be
\xi=\frac{y_{1}y_{2}}{t_{12}}\,.
\ee
The existence of the above cross ratio implies that the two-point function will be undetermined up to a function of the cross ratio $\xi$.

Next we solve the Ward identities for the rest of the symmetry generators. Requirement of the the boost invariance implies that the two-point function is~\footnote{In the present article, we focus mainly on operators with non-zero particle number. The situation with zero particle number is a subtle and we do not attempt to resolve it here. }
\be
<\mathcal T\mathcal O_{1}(t_{1},x_{1})\mathcal O_{2}(t_{2},x_{2})>=e^{-\frac{im_{\mathcal O_{1}}|\vec x_{12}|^{2}}{2t_{12}}}\delta_{m_{\mathcal O_{1}}+m_{\mathcal O_{2}},0}\,f(t_{12},y_{1},y_{2})\,.
\ee
Requirement of the scale invariance implies that the function $f(t,y_{1},y_{2})$ satisfies a differential equation given by
\be
\Big(2t_{12}\p_{t_{12}}+y_{1}\p_{y_{1}}+y_{2}\p_{y_{2}}+(\Delta_{\mathcal O_{1}}+\Delta_{\mathcal O_{2}})\Big)f(t_{12},y_{1},y_{2})=0\,.
\ee
The above equation states that the function $f(t,y_{1},y_{2})$ is a homogeneous function in variables $t_{12},y_{1}$ and $y_{2}$ with degree $-(\Delta_{\mathcal O_{1}}+\Delta_{\mathcal O_{2}})$. Thus, we write
\be
f(t_{12},y_{1},y_{2})=\frac{1}{t_{12}^{\frac{\Delta_{\mathcal O_{1}}+\Delta_{\mathcal O_{2}}}{2}}}g(\xi,\frac{y_{1}^{2}}{t_{12}},\frac{y_{2}^{2}}{t_{12}})\,.
\ee
Finally, solving the constraint coming from the invariance under the expansion transformation, we find
\be
f(t_{12},y_{1},y_{2})=\frac{c_{\mathcal O}\,\delta_{\Delta_{\mathcal O_{1}},\Delta_{\mathcal O_{2}}}}{t_{12}^{\frac{\Delta_{\mathcal O_{1}}+\Delta_{\mathcal O_{2}}}{2}}}e^{-\frac{im_{\mathcal O_{1}}(y_{1}-y_{2})^{2}}{2t_{12}}}G(\xi)\,,
\ee
where $c_{\mathcal O}$ is a constant.
Thus, the two-point function is 
\bea\label{TwoPointFn}
<\mathcal T\mathcal O_{1}(t_{1},x_{1})\mathcal O_{2}(t_{2},x_{2})>&=&\frac{c_{\mathcal O}\,\delta_{\Delta_{\mathcal O_{1}},\Delta_{\mathcal O_{2}}}\delta_{m_{\mathcal O_{1}}+m_{\mathcal O_{2}},0}}{t_{12}^{\frac{\Delta_{\mathcal O_{1}}+\Delta_{\mathcal O_{2}}}{2}}}e^{-\frac{im_{\mathcal O_{1}}x_{12}^{2}}{2t_{12}}}G(\xi)\nn\\
&=&\frac{c_{\mathcal O}\,\delta_{\Delta_{\mathcal O_{1}},\Delta_{\mathcal O_{2}}}\delta_{m_{\mathcal O_{1}}+m_{\mathcal O_{2}},0}}{(y_{1}y_{2})^{\frac{\Delta_{\mathcal O_{1}}+\Delta_{\mathcal O_{2}}}{2}}}e^{-\frac{im_{\mathcal O_{1}}x_{12}^{2}}{2t_{12}}}\tilde G(\xi)\,,
\eea
where
\be
x_{12}^{2}=(\vec x_{1}-\vec x_{2})^{2}+(y_{1}-y_{2})^{2}\,.
\ee
Note that the function $G(\xi)$ needs to satisfy a consistency condition that in the bulk limit, i.e. $\xi\rightarrow \infty$, the two-point function reduces to the bulk correlation function. This requires
\be
G(\xi)\rightarrow 1,\quad \text{as}\quad \xi\rightarrow\infty\,.
\ee
In the case of $d=1$, i.e. one spatial dimension in the presence of boundary at $y=0$, there is no Galilean symmetry. As a result, we do not have Ward identities corresponding to the translation, boost and $U(1)$ symmetry transformations. To obtain the one and two-point function, we need to solve the Ward identities for the $SL(2,\mathbb R)$ generators. 
Solving the Ward identities, we find that
\bea
&&<\mathcal O(t,y)>=\frac{c}{y^{\Delta_{\mathcal O}}}\,,\qquad\text{for}\quad m_{\mathcal O}=0\,,\nn\\
&&<\mathcal T\mathcal O_{1}(t_{1},y_{1})\mathcal O_{2}(t_{2},y_{2})>=\frac{c_{\mathcal O}\,\delta_{\Delta_{\mathcal O_{1}},\Delta_{\mathcal O_{2}}}}{t_{12}^{\frac{\Delta_{\mathcal O_{1}}+\Delta_{\mathcal O_{2}}}{2}}}e^{-\frac{i}{2t_{12}}\Big(m_{\mathcal O_{1}}y_{1}^{2}-m_{\mathcal O_{2}}y_{2}^{2}\Big)}G(\xi)\,.
\eea
The above result was first obtained in~\cite{Henkel:1993sg}.
\subsection{Boundary OPE} 
One of the characteristic features of both the relativistic and non-relativistic conformal field theory is the existence of a convergent operator product expansion (OPE). The OPE is the expansion of the product of two operators in terms of local operators and their descendants in a conformal field theory. 
In a relativistic conformal field theory in the presence of a boundary, there also exists a boundary OPE where a bulk primary operator near the boundary has an expansion in terms of the set of local boundary operators. 
In this section, we will find that a similar boundary OPE, with some differences, also exists in a non-relativistic conformal field theory. 

We consider a scalar primary operator, $\mathcal O_{1}(t,x)$, with non-zero particle number $m_{\mathcal O_{1}}$. The expansion of the field in terms of local boundary operators is restricted by the condition for the  $U(1)$ charge conservation, i.e. only those boundary operators appear in the expansion which has the particle number equals to $m_{\mathcal O_{1}}$. There is a further restriction coming from the conformal invariance of the two-point function between bulk and boundary operators.
Consider a scalar primary operator $\hat{\mathcal O}(t_{2},\vec x_{2})$ at the boundary. The two-point function between the bulk and the boundary scalar operator is
\be
<\mathcal T\mathcal O_{1}(t_{1},x_{1})\hat{\mathcal O}(t_{2},\vec x_{2})>=\frac{C^{b}_{\mathcal O_{1}\hat{\mathcal O}}\,\delta_{m_{\mathcal O_{1}}+m_{\hat{\mathcal O}}}}{y_{1}^{\Delta_{\mathcal O_{1}}-\Delta_{\hat{\mathcal O}}}t_{12}^{\Delta_{\hat{\mathcal O}}}}e^{-\frac{im_{\mathcal O_{1}}\vec x_{12}^{2}}{2t_{12}}}e^{-\frac{im_{\mathcal O_{1}}\,y_{1}^{2}}{2t_{12}}}\,.
\ee
Here $C^{b}_{\mathcal O_{1}\hat{\mathcal O}}$ is a constant.
Expanding the above two point function in powers of $y_{1}$, we get
\be\label{TwoPointFn.2}
<\mathcal T\mathcal O_{1}(t_{1},x_{1})\hat{\mathcal O}(t_{2},\vec x_{2})>=\frac{C^{b}_{\mathcal O_{1}\hat{\mathcal O}}\,\delta_{m_{\mathcal O_{1}}+m_{\hat{\mathcal O}}}}{t_{12}^{\Delta_{\hat{\mathcal O}}}}e^{-\frac{im_{\mathcal O_{1}}\vec x_{12}^{2}}{2t_{12}}}\sum_{n=0}^{\infty}\frac{1}{n!}\frac{1}{y_{1}^{\Delta_{\mathcal O_{1}}-\Delta_{\hat{\mathcal O}}}}\Big(-\frac{im_{\mathcal O_{1}}\,y_{1}^{2}}{2t_{12}}\Big)^{n}\,.
\ee
The above two point function suggests the following decomposition of the bulk scalar primary operator in terms of boundary primary operators
\be\label{boundaryOPE}
\mathcal O_{1}(t_{1},\vec x_{1},y_{1})=\sum_{\hat {\mathcal O}}\frac{B_{\mathcal O_{1}\,\hat{\mathcal O}}}{y_{1}^{\Delta_{\mathcal O_{1}}-\Delta_{\hat{\mathcal O}}}}\sum_{n=0}^{\infty}\frac{a_{n}}{n!}y_{1}^{2n}\mathcal D^{n}\hat{\mathcal O}(t_{1},\vec x_{1})\,,
\ee
where $\mathcal D$ is a differential operator of scaling dimension two that needs to be determined. Note that the boundary primary operator, $\hat{\mathcal O}$, that appears in the above expansion has the same particle number as the bulk primary operator. Substituting the boundary decomposition~\eqref{boundaryOPE} in the LHS of~\eqref{TwoPointFn.2} and using the two point function between boundary scalar primary operators
\be
<\mathcal T\hat{\mathcal O}_{1}(t_{1},\vec x_{1})\hat{\mathcal O}_{2}(t_{2},\vec x_{2})>=\frac{c_{\hat{\mathcal O}_{1}}\,\delta_{\Delta_{\hat{\mathcal O}_{1}},\Delta_{\hat{\mathcal O}_{2}}}\,\delta_{m_{\hat{\mathcal O}_{1}}+m_{\hat{\mathcal O}_{2}}}}{t_{12}^{\Delta_{\hat{\mathcal O}_{2}}}}e^{-\frac{im_{\hat{\mathcal O}_{1}}\vec x_{12}^{2}}{2t_{12}}}\,,
\ee
we obtain $C^{b}_{\mathcal O_{1}\hat{\mathcal O}}=B_{\mathcal O_{1}\,\hat{\mathcal O}}\,c_{\hat{\mathcal O}}$ and
\be
\mathcal D=\p^{2}_{i}-2im_{\mathcal O_{1}}\,\p_{t}\,,\quad\text{with}\quad a_{n}=\frac{(-1)^{n}}{2^{2n}\Big(\Delta_{\hat{\mathcal O}}-\frac{(d-1)}{2}\Big)_{n}}\,.
\ee
Here $d$ is the bulk spatial dimensions and $(x)_{n}$ is the Pochhammer symbol defined as
\be
(x)_{n}=x(x+1)(x+2)....(x+n-1)\,,
\ee
with $(x)_{0}=1$.

The boundary OPE has a particularly simpler form for the bulk primary operator with scaling dimension $\Delta=\frac{d}{2}$ and satisfying the free field equation
\be
(\p_{I}^{2}-2im_{\Phi}\p_{t})\Phi(t,x)=(\p_{y}^{2}+\p_{i}^{2}-2im_{\Phi}\p_{t})\Phi(t,x)=0\,.
\ee
In this case, the boundary OPE receives contributions only from the two primary operators of dimensions $\hat \Delta=\frac{d}{2}$ and $\hat \Delta=\frac{d}{2}+1$. We see this as follows: Imposing the free field equation to the boundary OPE, we obtain
\bea
&&(\p_{I}^{2}-2im_{\Phi}\p_{t})\Phi(t,x)=\sum_{\hat {\mathcal O}}\frac{B_{\Phi\,\hat{\mathcal O}}}{y^{\Delta_{\Phi}-\Delta_{\hat{\mathcal O}}}}\sum_{n=0}^{\infty}\frac{a_{n}}{n!}y^{2n}\mathcal D^{n+1}\hat{\mathcal O}(t,\vec x)\nn\\
&&\qquad\qquad\qquad+\sum_{\hat {\mathcal O}}\frac{B_{\Phi\,\hat{\mathcal O}}}{y^{\Delta_{\Phi}-\Delta_{\hat{\mathcal O}}}}\sum_{n=0}^{\infty}\frac{a_{n}}{n!}(2n-1+\hat\Delta-\frac{d}{2})(2n+\hat\Delta-\frac{d}{2})\,y^{2(n-1)}\mathcal D^{n}\hat{\mathcal O}(t,\vec x)\,,\nn\\
&&\qquad=\sum_{\hat {\mathcal O}}\frac{B_{\Phi\,\hat{\mathcal O}}}{y^{\Delta_{\Phi}-\Delta_{\hat{\mathcal O}}}}\sum_{n=1}^{\infty}\frac{a_{n-1}}{(n-1)!}y^{2n-2}\Big[1-\frac{(2n-1+\hat\Delta-\frac{d}{2})(2n+\hat\Delta-\frac{d}{2})}{4n(\hat\Delta-\frac{d}{2}+n-\frac{1}{2})}\Big]\mathcal D^{n}\hat{\mathcal O}(t,\vec x)\nn\\
&&\qquad+\sum_{\hat {\mathcal O}}\frac{B_{\Phi\,\hat{\mathcal O}}}{y^{2+\Delta_{\Phi}-\Delta_{\hat{\mathcal O}}}}(\hat\Delta-1-\frac{d}{2})(\hat\Delta-\frac{d}{2})\hat{\mathcal O}(t,\vec x)\,.
\eea
The above satisfies the free field equation provided the scaling dimensions are $\hat \Delta=\frac{d}{2}$ and $\hat \Delta=\frac{d}{2}+1$. Thus, the boundary OPE of a free scalar primary contains two boundary conformal multiplets corresponding to the boundary primary operators of dimensions $\hat \Delta=\frac{d}{2}$ and $\hat \Delta=\frac{d}{2}+1$. These primary operator corresponds to the boundary value of the bulk operator, i.e. $\Phi(t,\vec x,y=0)$ and its normal derivative, i.e. $\p_{y}\Phi(t,x)\Big|_{y=0}$.

It is straight forward to extend the above discussion to the $d=1$ case. Note that in this case, the boundary does not have any particle number symmetry, and as a result, the boundary primary operators are labelled by the scaling dimension only. To obtain the boundary OPE, the relevant two point functions are
\be
<\mathcal T\mathcal O_{1}(t_{1},y_{1})\hat{\mathcal O}(t_{2})>=\frac{\tilde C^{b}_{\mathcal O_{1}\hat{\mathcal O}}}{y_{1}^{\Delta_{\mathcal O_{1}}-\Delta_{\hat{\mathcal O}}}t_{12}^{\Delta_{\hat{\mathcal O}}}}e^{-\frac{im_{\mathcal O_{1}}\,y_{1}^{2}}{2t_{12}}}\,,\qquad <\mathcal T\hat{\mathcal O}_{1}(t_{1})\hat{\mathcal O}(t_{2})>=\frac{\tilde c_{\hat{\mathcal O_{1}}}\delta_{\hat\Delta_{1},\hat\Delta_{2}}}{t_{12}^{\hat\Delta_{2}}}.
\ee
Thus, the boundary decomposition is
\be\label{boundaryOPE.2}
\mathcal O_{1}(t_{1},y_{1})=\sum_{\hat {\mathcal O}}\frac{\tilde B_{\mathcal O_{1}\,\hat{\mathcal O}}}{y_{1}^{\Delta_{\mathcal O_{1}}-\Delta_{\hat{\mathcal O}}}}\sum_{n=0}^{\infty}\frac{\tilde a_{n}}{n!}y_{1}^{2n}(-2im_{1}\p_{t_{1}})^{n}\hat{\mathcal O}(t_{1})\,,
\ee
with
\be
\tilde a_{n}=\frac{(-1)^{n}}{2^{2n}\Big(\hat\Delta\Big)_{n}}\,,\quad\text{and}\quad \tilde C^{b}_{\mathcal O_{1}\hat{\mathcal O}}=\tilde B_{\mathcal O_{1}\,\hat{\mathcal O}}\,\tilde c_{\hat{\mathcal O}}\,.
\ee
\subsection{Boundary conformal block}
The boundary operator product expansion~\eqref{boundaryOPE} allows to decompose the bulk two point function in terms of two point functions of the boundary primary operators and their descendants.
It is possible to sum the contributions of the descendants for a given boundary primary operator. This will give rise the expansion of the bulk two point function in terms of boundary conformal block. To arrive at the boundary conformal block, we need 
\be
\mathcal D_{1}^{n}\mathcal D_{2}^{m}\frac{e^{-\frac{im_{\hat{\mathcal O}}\vec x_{12}^{2}}{2t_{12}}}}{t_{12}^{\Delta_{\hat{\mathcal O}}}}=(2im_{\hat{\mathcal O}})^{m+n}\Big(\Delta_{\hat{\mathcal O}}-\frac{d-1}{2}\Big)_{m+n}\,\frac{e^{-\frac{im_{\hat{\mathcal O}}\vec x_{12}^{2}}{2t_{12}}}}{t_{12}^{\Delta_{\hat{\mathcal O}}+m+n}}\,.
\ee
Using the above, the bulk two point function becomes
\bea
<\mathcal T\mathcal O_{1}(t_{1},x_{1})\mathcal O_{2}(t_{2},x_{2})>&=&\sum_{\hat {\mathcal O}}\frac{B_{\mathcal O_{1}\,\hat{\mathcal O}}B_{\mathcal O_{2}\hat{\mathcal O}}\,c_{\hat{\mathcal O}}\,\,\delta_{m_{\mathcal O_{1}}+m_{\mathcal O_{2}}}}{y_{1}^{\Delta_{\mathcal O_{1}}-\Delta_{\hat{\mathcal O}}}y_{2}^{\Delta_{\mathcal O_{2}}-\Delta_{\hat{\mathcal O}}}}\sum_{n,m=0}^{\infty}\frac{a_{n}a_{m}}{n!m!}y_{1}^{2n}y_{2}^{2m}(2im_{{\mathcal O_{1}}})^{m+n}\nn\\
&&\times\Big(\Delta_{\hat{\mathcal O}}-\frac{d-1}{2}\Big)_{m+n}\,\frac{e^{-\frac{im_{{\mathcal O_{1}}}\vec x_{12}^{2}}{2t_{12}}}}{t_{12}^{\Delta_{\hat{\mathcal O}}+m+n}}\,,\nn\\
&=&\frac{\delta_{m_{\mathcal O_{1}}+m_{\mathcal O_{2}},0}}{y_{1}^{\Delta_{\mathcal O_{1}}}y_{2}^{\Delta_{\mathcal O_{2}}}}\,e^{-\frac{im_{{\mathcal O_{1}}}\vec x_{12}^{2}}{2t_{12}}}\sum_{\hat {\mathcal O}}B_{\mathcal O_{1}\,\hat{\mathcal O}}B_{\mathcal O_{2}\hat{\mathcal O}}\,c_{\hat{\mathcal O}}\,\xi^{\Delta_{\hat{\mathcal O}}}\nn\\
&&\times\sum_{n,m=0}^{\infty}\frac{a_{n}a_{m}}{n!m!}\Big(\frac{y_{1}}{y_{2}}\Big)^{n-m}(2im_{{\mathcal O_{1}}}\xi)^{m+n}\Big(\Delta_{\hat{\mathcal O}}-\frac{d-1}{2}\Big)_{m+n}\,,\nn\\
&=&\frac{\delta_{m_{\mathcal O_{1}}+m_{\mathcal O_{2}},0}}{y_{1}^{\Delta_{\mathcal O_{1}}}y_{2}^{\Delta_{\mathcal O_{2}}}}\,e^{-\frac{im_{{\mathcal O_{1}}}x_{12}^{2}}{2t_{12}}}\sum_{\hat {\mathcal O}}B_{\mathcal O_{1}\,\hat{\mathcal O}}B_{\mathcal O_{2}\hat{\mathcal O}}\,c_{\hat{\mathcal O}}\,\xi^{\Delta_{\hat{\mathcal O}}}\nn\\
&&\times e^{-im_{{\mathcal O_{1}}}\xi}\,{}_{0}F_{1}\Big(\Delta_{\hat{\mathcal O}}-\frac{d-1}{2};-\frac{m_{{\mathcal O_{1}}}^{2}\xi^{2}}{4}\Big)\,.
\eea
In the last line, we have $x_{12}^{2}=\vec x_{12}^{2}+(y_{1}-y_{2})^{2}$. Comparing the above with~\eqref{TwoPointFn}, we will get the expression for the function $G(\xi)$ in terms of boundary conformal block.
\section{Non-relativistic conformal field theory on curved space-time with boundary}\label{NRCFTonCurvedSpace}
A quantum theory of fields coupled to the background metric and gauge fields contains rich information about the theory's spectrum of operators, phases, and symmetries. 
For example, the partition function of the theory on a curved spacetime as a function of background fields is used to derive Ward identities associated with symmetries and compute correlation functions of various conserved currents. In this direction, a particularly interesting case of investigation is the boundary conformal field theory, i.e. a conformal field theory on a manifold with a boundary. An important quantity to compute is the trace anomaly that provides a quantitative measure of the degrees of freedom in a conformal field theory. The trace anomaly for a conformal field theory on a curved manifold has expression in terms of geometric quantities constructed out of curvature tensor and Weyl tensor. The trace anomaly has a richer structure for a boundary conformal field theory, see~\cite{Herzog:2015ioa, Fursaev:2015wpa, Solodukhin:2015eca}. It receives contributions both from the bulk and boundary anomaly. An additional new structure appears that depends on the extrinsic curvature of the boundary. These bring new central charges that characterises the boundary conformal field theory, see for instance~\cite{Jensen:2015swa, Herzog:2017xha, Casini:2018nym, Kobayashi:2018lil, Wang:2021mdq}.

A $d$-dimensional Galilean invariant field theory can also be placed on a curved manifold. The necessary geometric structure required to couple a Galilean invariant field theory to the background is called the Newton-Cartan structure~\cite{Duval:1984cj, Duval:2009vt, Jensen:2014aia, Son:2005rv, Son:2013rqa, Geracie:2014nka}. The Newton-Cartan structure is specified by the providing the background fields $(n_{\m},A_{\m},h_{\m\n})$ where $n_{\m}$ defines the time direction, $A_{\m}$ couples to particle number/mass current and $h_{\m\n}$ is a spatial metric of rank $(d-1)$. It provides a natural set-up to define the energy-momentum tensor for a non-relativistic field theory. Therefore, an interesting question to answer in this direction is: What are the geometric structures that appear in the trace anomaly?

In this section, we construct a diffeomorphic and Weyl invariant action on a manifold with a boundary. A useful approach to constructing such a theory is by reducing a diffeomorphic and Weyl invariant relativistic field theory along a null isometry direction. As we will see below, this gives a diffeomorphic and Weyl invariant non-relativistic field theory on a manifold with boundary.

We will consider a theory of a massless complex scalar field in $(d+1)$-dimensions Lorentzian spacetime in the presence of co-dimension one boundary. The boundary is described by the embedding equation
 \be
 x^{M}=X^{M}(\sigma^{A})\,,
 \ee
 where $M=1,..,d+1$ and $A=1,...,d$. The coordinates $x^{M}$ and $\sigma^{A}$ are the bulk and boundary coordinates, respectively. Given the bulk metric $G=G_{MN}dx^{M}dx^{N}$, the induced metric at the boundary is
 \be
 \g_{AB}=G_{MN}e^{\m}_{M}e^{N}_{B},\quad \text{where}\quad e^{M}_{A}=\frac{\p X^{M}(\sigma^{A})}{\p \sigma^{A}}\,.
 \ee
 The diffeomorphic and Weyl invariant action is
 \bea
 S =-\frac{1}{4\pi}\int d^{d+1}x\sqrt{-G}\Big[G^{MN}\p_{M}\phi^{\dagger}\p_{N}\phi+\frac{d-1}{4d} R_{G}\phi^{\dagger}\phi\Big]-\frac{d-1}{8\pi d}\int_{\p\mathcal M}d^{d}\sigma\sqrt{-\gamma}\,K_{G}\phi^{\dagger}\phi\,.\nn\\
 \eea
In the above $R_{G}$ is the Ricci scalar of the bulk and $K_{G}$ is the trace of the extrinsic curvature of the boundary defined by
 \be
 K_{G,AB}=\frac{1}{2}\mathcal L_{\mathcal N}G_{MN}\,e^{M}_{A}e^{N}_{B}\,,
 \ee
 where $\mathcal N=\mathcal N^{M}\frac{\p}{\p x^{M}}$ is the vector normal to the boundary in the outward direction.  The above action is invariant under the Weyl transformation
 \be
 G\rightarrow e^{2\alpha(x)}G,\quad \phi\rightarrow e^{-\frac{d-1}{2}\alpha(x)}\phi\,.
 \ee 
 Next, we assume that the background metric has a null isometry, i.e. the metric has the following general~\cite{Jensen:2014aia}
 \be
 G=2n_{\m}dx^{\m}(dx^{-}+A_{\m}dx^{\m})+h_{\m\n}dx^{\m}dx^{\n}\,,\quad\text{where}\,\quad\m=1,...,d\,.
 \ee
In the above $x^{-}$ is the isometry direction and $n_{\m}$, $A_{\m}$ and $h_{\m\n}$ are independent of $x^{-}$. Note that in the above ansatz for the metric, we have incorporated the necessary geometric data, called Newton Cartan structure, to couple a non-relativistic conformal field theory to a curved background. The inverse of the metric is
\be
G^{-1}=(A^{2}-2v\cdot A)\p_{-}\otimes\p_{-}+2(v^{\m}-h^{\m\n}A_{\n})\p_{-}\otimes\p_{\n}+h^{\m\n}\p_{\m}\otimes\p_{\n}\,.
\ee
In the above we have introduced tensor fields $v^{\m}$ and $h^{\m\n}$ which are defined as
\be
v^{\m}n_{\m}=1,\quad h_{\m\n}v^{\n}=0,\quad h^{\m\n}n_{\n}=0,\quad h^{\m\n}h_{\n\rho}=\delta^{\m}_{\rho}-v^{\m}n_{\rho}\,.
\ee
The scalar quantities are
\be
A^{2}=h^{\m\n}A_{\m}A_{\n},\quad \text{and}\quad v\cdot A=v^{\m}A_{\m}\,.
\ee
Next, we compactify the null direction $x^{-}$ with periodicity $2\pi$. Writing the field $\phi$ as
 \be
 \phi(x^{-},x^{\m})=e^{imx^{-}}\psi(x^{\m})\,,
 \ee
 the reduced action is
 \bea\label{ActionwithBoundary}
 S &=&-\frac{1}{2}\int d^{d}x\sqrt{\text{det}(n_{\m}n_{\n}+h_{\m\n})}\Big[(A^{2}-2v\cdot A)m^{2}\psi\psi^{\dagger}-im\,(v^{\m}-h^{\m\n}A_{\n})(\psi^{\dagger}\p_{\m}\psi-\p_{\m}\psi^{\dagger}\psi)\nn\\
 &&\quad\quad\quad\quad\quad\quad+h^{\m\n}\p_{\m}\psi^{\dagger}\p_{\n}\psi+\frac{d-1}{4d} R_{G}\,\psi^{\dagger}\psi\Big]
-\frac{d-1}{4d}\int_{\p\mathcal M}d^{d-1}\sigma\sqrt{-\gamma}\,K_{G}\,\psi^{\dagger}\psi\,,\nn\\
&=&-\frac{1}{2}\int d^{d}x\sqrt{\text{det}(n_{\m}n_{\n}+h_{\m\n})}\Big[-im\,v^{\m}(\psi^{\dagger}D_{\m}\psi-D_{\m}\psi^{\dagger}\psi)+h^{\m\n}D_{\m}\psi^{\dagger}D_{\n}\psi\nn\\
&&\quad\quad\quad\quad\quad\quad\qquad\qquad+\frac{d-1}{4d} R_{G}\,\psi^{\dagger}\psi\Big]
-\frac{d-1}{4d}\int_{\p\mathcal M}d^{d-1}\sigma\sqrt{-\gamma}\,K_{G}\,\psi^{\dagger}\psi\,.
 \eea 
where the gauge covariant derivative is given by
 \be
 D_{\m}\psi=(\p_{\m}-imA_{\m})\psi\,.
 \ee
The above action is invariant under the general coordinate transformations $x^{\m}\rightarrow x'^{\m}$ and the Weyl transformation
 \be
 n_{\m}\rightarrow e^{2\alpha(x)}n_{\m},\quad A_{\m}\rightarrow A_{\m},\quad h_{\m\n}\rightarrow e^{2\alpha(x)}h_{\m\n},\quad \psi\rightarrow e^{-\frac{d-1}{2}\alpha(x)}\psi\,.
 \ee
It is important to note that in the action~\eqref{ActionwithBoundary}, the Ricci scalar and the extrinsic curvature are constructed from the original metric $G$. Generalising the above action~\eqref{ActionwithBoundary} to include interactions localized at the boundary, we have 
\bea\label{ActionwithBoundary2}
S &=&-\frac{1}{2}\int d^{d}x\sqrt{\text{det}(n_{\m}n_{\n}+h_{\m\n})}\Big[-im\,v^{\m}(\psi^{\dagger}D_{\m}\psi-D_{\m}\psi^{\dagger}\psi)+h^{\m\n}D_{\m}\psi^{\dagger}D_{\n}\psi\nn\\
&&\quad+\frac{d-1}{4d} R_{G}\,\psi^{\dagger}\psi\Big]
-\frac{d-1}{4d}\int_{\p\mathcal M}d^{d-1}x\sqrt{-\gamma}\,K_{G}\,\psi^{\dagger}\psi-\frac{1}{2}\int_{\p\mathcal M}d^{d-1}\sigma\sqrt{-\gamma}\,\mathcal L_{\text{Int}}\,.\nn\\
\eea
The flat space limit is obtained by substituting
\be
n=dt,\quad A_{\m}=0,\quad h_{\m\n}=\delta_{ij}\,,
\ee
where $\m=(t,i)$ and $i=1,...,(d-1)$. In the subsequent section, we will be interested in non-relativistic field theories in the flat space with a co-dimension one planar boundary located at $y=0$, where $y$ is one of the spatial coordinates.
In this case, the action becomes
\be
S=-\frac{1}{2}\int dt\,d^{d-1}x\Big[-im\,(\psi^{\dagger}\p_{t}\psi-\p_{t}\psi^{\dagger}\psi)+\delta^{ij}\p_{i}\psi^{\dagger}\p_{j}\psi\Big]-\frac{1}{2}\int_{y=0}dt\,d^{d-2}\vec x\,\mathcal L_{\text{Int}}\,,
\ee
where $x\equiv (y,\vec x)$ and $\vec x$ denotes the boundary coordinates.
The equation of motion and the boundary conditions are
\be
2im\,\p_{t}\psi+\delta^{ij}\p_{i}\p_{j}\psi=0,\quad \mathcal N^{i}\p_{i}\psi-\frac{\delta\mathcal L_{\text{Int}}}{\delta\psi^{\dagger}}=0\,,
\ee
where $N$ is the vector normal to the boundary in the outward direction.

For a conformal field theory in the presence of a boundary, there exists a special boundary operator, called the displacement operator, whose scaling dimension is the same as the energy-momentum tensor. The displacement operator plays an important role in constraining the boundary anomalies in various dimensions~\cite{Herzog:2017kkj}. 
With similar motivation, starting with the action~\eqref{ActionwithBoundary2}, one can derive an expression for a boundary operator, which can be regarded as a non-relativistic analogue of the displacement operator.  Restricting ourselves to the flat space and the planar boundary at $y=0$, if we define the displacement operator as the change in the action due to the displacement of the boundary by an infinitesimal amount $\delta y(\vec x)$ as
\be
\delta_{y} S=-\int_{y=0}dt\,d^{d-2}\vec x\,\delta y(\vec x)\,D(t,\vec x)\,,
\ee
then we obtain the following expression for the displacement operator in terms of the bulk fields 
\be
D(t,\vec x)=(\p_{y}\psi^{\dagger})(\p_{y}\psi)-\frac{1}{4d}\delta^{ij}\p_{i}\p_{j}(\psi^{\dagger}\psi)-\frac{d-1}{4d}\p_{y}^{2}(\psi^{\dagger}\psi)\,.
\ee
It would be interesting to investigate the boundary anomalies in a non-relativistic conformal field theory and the role of the displacement operator in constraining these anomalies.
\section{Free fermions with boundary interactions}\label{Freefermions}
In this section, we consider non-relativistic theories of free fermions with boundary interactions and find the fixed point of the renormalization group flow. At the fixed point, the theory will be described by a non-relativistic conformal field theory. We compute the scaling dimension of some of the boundary primary operators. In Appendix~\ref{SigmaPhiTheory}, we also present an example of non-relativistic $\sigma\phi$-theory with boundary interaction and find the fixed point.
\subsection{Fermion interacting with a complex scalar at boundary}
We consider a 2-component free fermion in the bulk. We also introduce a complex scalar field at the boundary. The two degrees of freedoms are interacting with each other with interactions localized at the boundary. The action is
\bea
S&=&\int dt\,d^{d}x\, (i\psi_{\sigma}^{\dagger}\p_{t}\psi_{\sigma}-\frac{1}{2}\p_{\m}\psi_{\sigma}^{\dagger}\p_{\m}\psi_{\sigma})+\int dt\,d^{d-1}x\,\Big[g\,\psi^{\dagger}_{\uparrow}\psi^{\dagger}_{\downarrow}\phi+g\,\psi_{\downarrow}\psi_{\uparrow}\phi^{*}\Big]\nn\\
&&+\int dt\,d^{d-1}x\, \Big[i\phi^{*}\p_{t}\phi-\frac{1}{4}\p_{i}\phi^{*}\p_{i}\phi\Big]+\lambda\int dt\,d^{d-1}x\, (\phi^{*}\phi)^{2}\,.
\eea
We can rewrite the fermionic part of the above action in the matrix form as
\bea
S&=&\int dt\,d^{d}x\, \Psi^{\dagger}\Big(i\p_{t}+\sigma_{3}\frac{\nabla^{2}}{2}\Big)\Psi+\int dt\,d^{d-1}x\,\Big[g\,\Psi^{\dagger}\sigma_{-}\Psi\,\phi^{*}+g\,\Psi^{\dagger}\sigma_{+}\Psi\,\phi\Big]\nn\\
&&+\int dt\,d^{d-1}x\, \Big[i\phi^{*}\p_{t}\phi-\frac{1}{4}\p_{i}\phi^{*}\p_{i}\phi\Big]+\lambda\int dt\,d^{d-1}x\, (\phi^{*}\phi)^{2}\,,
\eea
where $\Psi=(\psi_{\uparrow},\psi_{\downarrow}^{\dagger})^{T}$ and $\sigma_{3,\pm}$ are Pauli matrices.
Dimensional analysis suggests that the coupling is marginal in $d=3$ and relevant for $d<3$. Therefore, to perform the $\e$-expansion, we will work in the dimensions $d=3-\e$.

\begin{figure}[htpb]
\begin{center}
\vspace{-3.5 cm}
\centering
\includegraphics[width=8in]{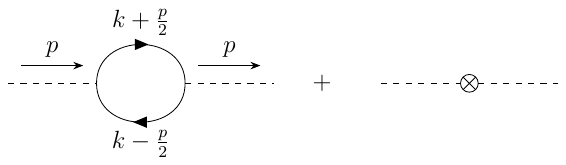}
\vspace{-20cm}
\caption{Self energy calculation at one-loop order. The solid (dashed) line represents fermion (scalar) propagator.\label{2-pointFn1}}
\end{center}
\end{figure}

Fermions are free in the bulk, and therefore, one does not expect them to acquire a wave function renormalization. The scalar field on the other hand acquires the wave function renormalization. This results in the renormalization of the coupling constant. 
We compute the wave function renormalization of the scalar field at one loop order. The relevant diagrams are shown in figure. \ref{2-pointFn1}. The contribution of the fermionic one loop is
\be
-2\frac{1}{2}(ig)^{2}\int\frac{dE\,d^{d-1}\vec k}{(2\pi)^{d}}G^{\p\p}_{11}(E+\frac{p^{0}}{2},\vec k+\frac{\vec p}{2})G^{\p\p}_{22}(E-\frac{p^{0}}{2},\vec k-\frac{\vec p}{2})\,.
\ee
After performing the contour integrating over $E$, the above contribution becomes
\bea
&&-4ig^{2}\int\frac{d^{d-1}\vec k\,dk_{y}dk'_{y}}{(2\pi)^{d+1}}\frac{1}{p^{0}-\frac{(\vec k+\frac{\vec p}{2})^{2}}{2}-\frac{(\vec k-\frac{\vec p}{2})^{2}}{2}-\frac{k^{2}_{y}+k^{'2}_{y}}{2}+i\delta}\nn\\
&&=-8ig^{2}S_{d}\int^{\infty}_{0}\frac{k^{d}dk}{(2\pi)^{d+1}}\frac{1}{p^{0}-2\frac{\vec k^{2}}{2}-\frac{\vec p^{2}}{4}+i\delta}=\frac{4\pi i}{(2\pi)^{d+1}\cos\frac{\pi d}{2}}g^{2}S_{d}\,(-p^{0}+\frac{\vec p^{2}}{4})^{\frac{d-1}{2}}\,.
\eea
Now, using the relations
\be
\frac{1}{\cos\frac{\pi d}{2}}\Big|_{d=3-\e}=-\frac{2}{\pi\e}+\mathcal O(\e),\quad \text{and}\quad S_{d}=\frac{2\pi^{\frac{d+1}{2}}}{\Gamma(\frac{d+1}{2})}\,,
\ee
we obtain the divergent contribution of the fermionic loop to be
\be
-\frac{ig^{2}}{\pi^{2}\e}(-p^{0}+\frac{\vec p^{2}}{4})\,.
\ee
The contribution of the counter term in the two-point function is
\be
i\delta_{\phi}(p^{0}-\frac{\vec p^{2}}{4})\,,
\ee 
where $Z_{\phi}=1+\delta_{\phi}$.

\begin{figure}[htpb]
\begin{center}
\vspace{-3.5 cm}
\centering
\includegraphics[width=8in]{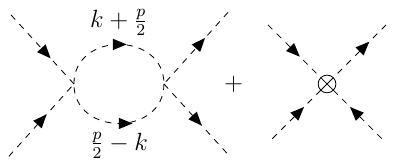}
\vspace{-18cm}
\caption{Four point function at one loop.\label{4ptFunction2}}
\end{center}
\end{figure}

Requiring that the residue at the pole, $p^{0}-\frac{\vec p^{2}}{4}=0$, is $1$, we obtain
\be
\delta_{\phi}=-\frac{g^{2}}{\pi^{2}\e}\,.
\ee
To renormalize the coupling constant, we also need to introduce the counter term $\delta_{g}$ as
\be
g\sqrt{Z_{\phi}}=(g_{R}+\delta_{g})\m^{\frac{\e}{2}}\,.
\ee
However, the three-point function involving $\Psi^{\dagger},\Psi$ and $\phi$ does not have any non trivial one particle irreducible diagrams at one loop order. Thus, the running of the coupling constant $g$ is governed by the wave function renormalization $Z_{\phi}$. To determine the renormalization of the vertex $(\phi^{*}\phi)^{2}$, we need to compute the four-point function. The relevant diagrams are shown in the figure \ref{4ptFunction2}. The Feynman diagram with fermionic loop trivially vanishes.
The contribution due to the loop and counter term are
\be
16\frac{1}{2}(i\lambda_{R})^{2}\int\frac{dE\,d^{d-1}k}{(2\pi)^{d}}D(k+\frac{p_{s}}{2})D(-k+\frac{p_{s}}{2})+4i\m^{\e}\lambda_{R}\,\delta_{\lambda}\,.
\ee
Here the scalar propagator is
\be
D(k)=\frac{i}{k^{0}-\frac{\vec k^{2}}{4}+i\delta}\,.
\ee
and the renormalized coupling is defined using the relation
\be
\lambda Z^{2}_{\phi}=\m^{\e}\lambda_{R}\,Z_{\lambda}=\m^{\e}\lambda_{R}\,(1+\delta_{\lambda})\,.
\ee
Evaluating the above in $d=3-\e$, we obtain the divergent contribution to be
\be
\frac{8i\lambda_{R}^{2}}{2\pi }\frac{\pi}{\cos\frac{\pi}{2}(1-\e)}+4i\m^{\e}\lambda_{R}\delta_{\lambda}=\frac{8i\lambda_{R}^{2}}{\pi\e}+4i\m^{\e}\lambda_{R}\delta_{\lambda}\,.
\ee
Cancelling the divergence, we obtain
\be
\delta_{\lambda}=-\frac{2\lambda_{R}}{\pi\e}\,.
\ee
Finally, computing the beta functions, we get
\be
\beta_{g}=-\frac{\e}{2}g_{R}+\frac{ g_{R}^{3}}{2\pi^{2}}\,,\quad \beta_{\lambda}=-\e \lambda_{R}-\frac{2\lambda_{R}^{2}}{\pi}+\frac{2g_{R}^{2}\lambda_{R}}{\pi^{2}}\,.
\ee
We see that there are two interacting IR fixed points
\bea
(g^{2}_{R},\lambda_{R})=(\pi^{2}\e,0),\quad\text{and}\quad (\pi^{2}\e,\frac{\pi\e}{2})\,.
\eea
At these fixed points, the theory is described by a non-relativisitic conformal field theory.

\begin{figure}[htpb]
\begin{center}
\vspace{-3 cm}
\centering
\includegraphics[width=8in]{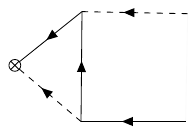}
\vspace{-19cm}
\caption{One loop computation of scaling dimension of three-fermion boundary operator.\label{ThreeFermionOp1}}
\end{center}
\end{figure}

\begin{figure}[htpb]
\begin{center}
\vspace{-3 cm}
\centering
\includegraphics[width=8in]{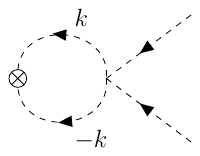}
\vspace{-19cm}
\caption{One loop computation of scaling dimension of four-fermion boundary operator.\label{4FermionOp1}}
\end{center}
\end{figure}

Next, we compute the scaling dimension of some of the boundary primary operators. The lowest dimension operator with fermionic particle number equals to one is $\psi_{\uparrow,\downarrow}\Big|_{y=0}$. Its scaling dimension is 
\be
\Delta_{\psi}=\frac{d}{2}=\frac{3}{2}-\frac{\e}{2}\,.
\ee
The lowest dimension operator with two-fermion particle number is the scalar field $\phi$. The anomalous dimension of the scalar field at the fixed point is
\be
\g_{\phi}=\frac{\m}{2}\frac{d}{d\m}\ln Z_{\phi}=\frac{g_{R}^{2}}{2\pi^{2}}=\frac{\e}{2}\,.
\ee
Thus, the scaling dimension of the two-fermion boundary operator $\phi$ is
\be
\Delta_{\phi}=\frac{d-1}{2}+\frac{\e}{2}=1\,.
\ee 
Next, we calculate the scaling dimension of three-fermion boundary operator. The simplest operator with zero angular momentum is $\phi\psi_{\uparrow}$. It is the lowest dimension operator with three fermion particle numbers. For the computation of its scaling dimension at one loop, we need to evaluate the diagram shown in the figure~\ref{ThreeFermionOp1}\footnote{Note that there is an additional contribution due to the digram, where the external scalar propagator is the one-loop corrected propagator. 
We have not shown the diagram; however, we will take its contribution into account while calculating the scaling dimension of the composite operators.}.
The contribution is
\be
-g_{R}^{2}\int\frac{dE\,d^{d-1}\vec k}{(2\pi)^{d}}G_{11}(k)G_{22}(k)D(-k)\,.
\ee
Explicit computation shows that the above integral is finite and therefore, the one loop contribution to the anomalous dimension vanishes (more precisely, it is non trivial part, there is also a trivial contribution due to the anomalous dimension of $\phi$). Thus, the scaling dimension of the three fermion boundary operator is
\be
\Delta_{\phi\psi_{\uparrow}}=\Delta_{\phi}+\Delta_{\psi_{\uparrow}}=1+\frac{d}{2}=\frac{5}{2}-\frac{\e}{2}\,.
\ee
Let us calculate the dimension of $\phi^{2}$ operator. This is the operator with fermion number 4. The leading contribution to the anomalous dimension comes from the one loop diagram shown in the figure~\ref{4FermionOp1}.
The contribution is
\be
2(1+\delta_{\phi^{2}})+4(i\lambda)\int \frac{dE\,d^{d-1}\vec k}{(2\pi)^{d}}D(k)D(-k)
\ee
Cancellation of the divergence requires that
\be
\delta_{\phi^{2}}=-\frac{2\lambda_{R}}{\pi\e}\,.
\ee
Therefore, the anomalous dimension at one loop is
\be
\gamma_{\phi^{2}}=-\frac{2\lambda_{R}}{\pi}\,.
\ee
We note that the anomalous dimension vanishes at the fixed point $(\pi^{2}\e,0)$. 
Thus, we have the following scaling dimensions at the one loop order
\bea
&&\text {At the fixed point $(\pi^{2}\e,0)$}: \Delta_{\phi^{2}}=2\Delta_{\phi}=2\,,\nn\\
&&\text {At the fixed point $\Big(\pi^{2}\e,\frac{\pi\e}{2}\Big)$}: \Delta_{\phi^{2}}=2+\gamma_{\phi^{2}}=2-\e\,.
\eea

\subsection{Fermions with quartic interaction}
\begin{figure}[htpb]
\begin{center}
\vspace{-3.5 cm}
\centering
\includegraphics[width=8in]{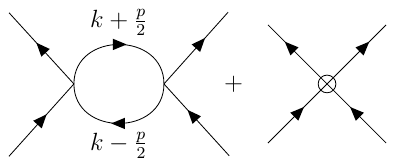}
\vspace{-18cm}
\caption{Four point function at one loop.\label{4PtFunctionFermion}}
\end{center}
\end{figure}

\begin{figure}[htpb]
\begin{center}
\vspace{-3 cm}
\centering
\includegraphics[width=8in]{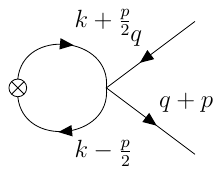}
\vspace{-19cm}
\caption{One loop computation of scaling dimension of two-fermion boundary operator.\label{2-fermionBoundaryOp}}
\end{center}
\end{figure}

\bea
S&=&\int dt\,d^{d}x\, (i\psi_{\sigma}^{\dagger}\p_{t}\psi_{\sigma}-\frac{1}{2}\p_{\m}\psi_{\sigma}^{\dagger}\p_{\m}\psi_{\sigma})+\int dt\,d^{d-1}x\,\Big[g\,\psi^{\dagger}_{\uparrow}\psi^{\dagger}_{\downarrow}\psi_{\downarrow}\psi_{\uparrow}\Big]\,,\nn\\
&=&\int dt\,d^{d}x\, \Psi^{\dagger}\Big(i\p_{t}+\sigma_{3}\frac{\nabla^{2}}{2}\Big)\Psi+\int dt\,d^{d-1}x\,\Big[g\,(\Psi^{\dagger}\sigma_{-}\Psi)\,(\Psi^{\dagger}\sigma_{+}\Psi)\Big]\,.
\eea
The coupling constant $g$ is marginal in the dimension $d=1$. We will work in the dimensions $d=1+\e$. 
We would like to emphasise here that in the dimension $d=1$, we do not have the particle number symmetry at the boundary. But at the end of the day, we will put $\e=1$, therefore, we will classify the particle based on the particle number symmetry. The relevant diagrams in the computation of the beta function at one loop are shown in the figure~\ref{4PtFunctionFermion}.
The contributions of these diagrams are
\be
-2\frac{1}{2}(ig_{R})^{2}\int\frac{dE\,d^{d-1}\vec k}{(2\pi)^{d}}G^{\p\p}_{11}(E+\frac{p^{0}}{2},\vec k+\frac{\vec p}{2})G^{\p\p}_{22}(E-\frac{p^{0}}{2},\vec k-\frac{\vec p}{2})+ig_{R}\delta_{g}\,,
\ee
where the renormalized coupling is defined to be
\be
g=g_{R}\m^{-\e}(1+\delta_{g})\,.
\ee
Cancelling the divergence, we obtain
\be
\delta_{g}=\frac{4g_{R}}{\pi\e}\,.
\ee
The corresponding $\beta$-function is
\be
\beta=\e\,g_{R}-\frac{4g_{R}^{2}}{\pi}\,.
\ee
Thus, it has a non-trivial fixed point at
\be
g^{*}_{R}=\frac{\pi\e}{4}\,.
\ee
Next, we compute the scaling dimension of some of the boundary primary operators. The lowest dimension operator with fermionic particle number equals to one is $\psi_{\uparrow,\downarrow}\Big|_{y=0}$. Its scaling dimension is 
\be
\Delta_{\psi}=\frac{d}{2}=\frac{1}{2}+\frac{\e}{2}\,.
\ee
Next, we calculate the scaling dimension of the two fermion boundary operator $\psi_{\downarrow}\psi_{\uparrow}=\Psi^{\dagger}\sigma_{-}\Psi$. The relevant Feynman diagram is shown in the figure~\ref{2-fermionBoundaryOp}.
\be
-(1+\delta_{\psi_{\downarrow}\psi_{\uparrow}})+ig\int\frac{dEd^{d-1}\vec k}{(2\pi)^{d}}G^{\p\p}_{11}(E+\frac{p_{0}}{2},\vec k+\frac{\vec p}{2})G^{\p\p}_{22}(E-\frac{p_{0}}{2},\vec k-\frac{\vec p}{2})
\ee
Cancellation of divergence implies that
\be
\delta_{\psi_{\downarrow}\psi_{\uparrow}}=\frac{4g_{R}}{\pi\e}\,.
\ee
Thus, the anomalous and the scaling dimension of the 2-fermion boundary operator are
\be
\g_{\psi_{\downarrow}\psi_{\uparrow}}=-\frac{4g^{*}_{R}}{\pi}=-\e,\quad\Delta_{\psi_{\downarrow}\psi_{\uparrow}}=d-\e=1\,.
\ee
We can continue this to compute the scaling dimension of various boundary primary operators.
\section{Discussion}
In this paper, we have initiated the study of Schr$\ddot{\text{o}}$dinger invariant field theory on a manifold with a boundary. 
The presence of the boundary breaks the Schr$\ddot{\text{o}}$dinger symmetry of the bulk to the subgroup of transformations which preserve the boundary.
In the case of flat space with a planar boundary, the reduced group includes the centrally extended Galilei group of the boundary, dilatations and expansion transformation. The primary operators are, therefore, labeled by angular momentum in the plane of the boundary, particle number and scaling dimension. 

The invariance of the correlation function under the reduced Schr$\ddot{\text{o}}$dinger group gives rise to the necessary Ward identities.
We solved these Ward identities to obtain the expression for the one and two-point functions of the bulk scalar primary operators with non-zero particle numbers. The correlation function exists when the total particle number of the product of the operators vanishes. As in a relativistic CFTs with boundary, the two-point function is fixed up to a function of a Schr$\ddot{\text{o}}$dinger invariant cross-ratio. An interesting aspect of conformal field theory in the presence of boundary is that there is another set of conformal multiplets, which are localized on the boundary. Using the conformal invariance and the correlation function between bulk and boundary scalar primary operators, we have obtained the expression for the boundary OPE.  
A bulk primary operator with a given particle number in the spatial dimensions $d>1$ decomposes into the sum of boundary conformal multiplets with various scaling dimension but with the same particle number. However, in the dimension $d=1$, there is no constraint of particle number conservation.
Furthermore, summing the contributions from a given conformal multiplet, we are able to obtain the expression for the boundary conformal block. 
We also discussed the diffeomorphic and Weyl invariant action for the Schr$\ddot{\text{o}}$dinger field on a manifold with a boundary. This requires the coupling of the Schr$\ddot{\text{o}}$dinger field to the Newton-Cartan structure on the orientable manifold. It provides a natural set-up to define the energy-momentum tensor for a non-relativistic field theory in the presence of the boundary. 

Finally, we conclude by highlighting some of the problems left for future work.
\begin{enumerate}
\item In the paper, we have not addressed the issue of the two-point correlation function of operators with the zero particle number. As a result, we have missed the correlation functions of some of the operators of interests like energy-momentum tensor. The zero particle sector in Schr$\ddot{\text{o}}$dinger field theory is subtle and requires care, as have been explained in~\cite{Golkar:2014mwa}.
\item The energy-momentum tensor of a Schr$\ddot{\text{o}}$dinger invariant field theory on flat space satisfies the trace identity
\be
2T^{0}_{0}+T^{i}_{i}=0\,.
\ee
The above identity modifies in the presence of the non-trivial Newton-Cartan background. This gives rise to non-relativistic scale anomalies~\cite{Jensen:2014hqa, Arav:2016xjc, Auzzi:2015fgg, Auzzi:2017wwc}. It would be interesting to extend the analysis in the presence of a boundary. As in a relativistic boundary conformal field theory, one would expect that the presence of the boundary introduces new geometric structures in the trace identity. The coefficients of these terms would give rise to new central charges.
\item The monotonic behaviour of the renormalization group flow in a relativistic quantum field theory has been investigated quite extensively. It would be interesting to carry out a similar investigation in the case of non-relativistic quantum field theory. More precisely, it would be interesting to associate a quantitative measure, like central charge, which decreases along with the RG flow. 
\item Finally, studying the supersymmetric version of Galilean field theory would be interesting. It would be nice to see a generalization of the idea of supersymmetric localization to non-relativistic field theories. Then one can use the tool of the supersymmetric localization to compute the partition function of a strongly interacting Schr$\ddot{\text{o}}$dinger theory on a manifold with and without boundary.
\end{enumerate}
\section*{Acknowledgments}
 This work is supported by the ISIRD grant 9-406/2019/IITRPR/5480. We would like to thank Shankhadeep Chakrabortty and Christopher Herzog for the useful discussion.

\appendix
\section{Feynman Rules}
We write down the fermionic propagator for the theory
\bea
S&=&\int dt\,d^{d}x\, \Psi^{\dagger}\Big(i\p_{t}+\sigma_{3}\frac{\nabla^{2}}{2m}\Big)\Psi+\int dt\,d^{d-1}x\,\mathcal L_{{\text{Int.}}}\,.
\eea
The bulk to bulk propagator for fermion with Neumann boundary condition is given by
\be
<\mathcal T\Psi(t,\vec x,y)\Psi^{\dagger}(t',\vec x',y')>=\int\frac{dE\,d^{d-1}\vec p}{(2\pi)^{d}}e^{-iE(t-t')-i\vec p\cdot(\vec x-\vec x')}G(E,\vec p ; y,y')\,,
\ee
where
\be
G(E,\vec p ; y,y')=\int_{-\infty}^{\infty}\frac{dp_{y}}{2\pi}\frac{i}{E^{2}-\e_{ p}^{2}+i\delta}\begin{pmatrix}E+\e_{p}&&0\\
0&&E-\e_{p}\end{pmatrix}(e^{-ip_{y}(y-y')}+e^{-ip_{y}(y+y')})\,.
\ee
Here
\be
\e_{p}=\e_{\vec p}+\frac{p_{y}^{2}}{2m},\quad\text{where}\quad \e_{\vec p}=\frac{\vec p^{2}}{2m}\,.
\ee
In particular, the boundary to boundary propagator is
\be
G^{\p\p}(E,\vec p)=2\int_{-\infty}^{\infty}\frac{dp_{y}}{2\pi}\frac{i}{E^{2}-\e_{ p}^{2}+i\delta}\begin{pmatrix}E+\e_{p}&&0\\
0&&E-\e_{p}\end{pmatrix}\,.
\ee

\section{Non-relativistic $\sigma\phi$-theory}\label{SigmaPhiTheory}
\begin{figure}[htpb]
\begin{center}
\vspace{-3.5 cm}
\centering
\includegraphics[width=8in]{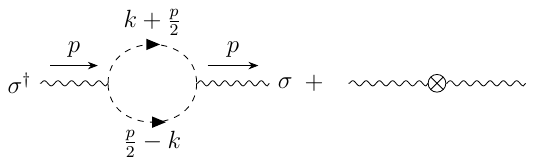}
\vspace{-20cm}
\caption{Wavy line represents $\sigma$ propagator and the dashed line represents $\phi^{I}$ propagator.\label{2-pointFn2}}
\end{center}
\end{figure}

\begin{figure}[htpb]
\begin{center}
\vspace{-3 cm}
\centering
\hspace*{-3.5cm} 
\includegraphics[width=8in]{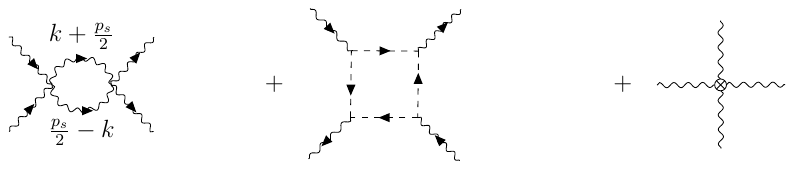}
\vspace{-18cm}
\caption{Wavy line represents $\sigma$ propagator and the dashed line represents $\phi^{I}$ propagator.\label{4-pointFn}}
\end{center}
\end{figure}

Consider the action
\bea
S&=&\int dt\,d^{d}x\, \Big[i\phi^{I\dagger}\p_{t}\phi^{I}-\frac{1}{2m_{1}}\p_{\m}\phi^{I\dagger}\p_{\m}\phi^{I}\Big]+\int dt\,d^{d-1}x\, \Big[i\sigma^{\dagger}\p_{t}\sigma-\frac{1}{2m_{2}}\p_{\m}\sigma^{\dagger}\p_{\m}\sigma\nn\\
&&+\lambda_{1}\sigma\phi^{I\dagger}\phi^{I\dagger}+\lambda_{1}\sigma^{\dagger}(\phi^{I}\phi^{I})+\lambda_{2}(\sigma^{\dagger}\sigma)^{2}\Big]\,.
\eea
Here $I=1,...,N$ and $\phi^{I}$ and $\sigma$ are scalars. Here $m_{2}=2m_{1}$. We will work in $d=3-\e$. In this case, the mass dimensions of $\lambda_{1}$ is $\frac{\e}{2}$ and the dimension of $\lambda_{2}$ is $\e$.\\
We start with the self energy calculation for the $\sigma$ field. The relevant diagrams at one loop are shown in figure \ref{2-pointFn2}. The contribution due to $\phi^{I}$ loop is
\be
2\frac{1}{2}2N(i\lambda_{1})^{2}\int\frac{dE\,d^{d-1}k}{(2\pi)^{d}}D^{\p\p}(k+\frac{p}{2})D^{\p\p}(\frac{p}{2}-k)\,.
\ee
Calculating the above, we get
\be
8N\lambda_{1}^{2}\int\frac{dE\,d^{d-1}k}{(2\pi)^{d}}\frac{dk_{y}\,dk'_{y}}{(2\pi)^{2}}\frac{1}{(E+\frac{p^{0}}{2}-\frac{(\vec k+\frac{\vec p}{2})^{2}}{2m_{1}}-\frac{k^{2}_{y}}{2m_{1}}+i\delta)(\frac{p^{0}}{2}-E-\frac{(-\vec k+\frac{\vec p}{2})^{2}}{2m_{1}}-\frac{k^{'2}_{y}}{2m_{1}}+i\delta)}\,.
\ee
Picking the pole in the upper half plane, we obtain
\be
-8iN\lambda_{1}^{2}\int\frac{d^{d-1}k}{(2\pi)^{d-1}}\frac{dk_{y}\,dk'_{y}}{(2\pi)^{2}}\frac{1}{(p^{0}-\frac{(\vec k+\frac{\vec p}{2})^{2}}{2m_{1}}-\frac{k^{2}_{y}}{2m_{1}}-\frac{(-\vec k+\frac{\vec p}{2})^{2}}{2m_{1}}-\frac{k^{'2}_{y}}{2m_{1}}+i\delta)}\,.
\ee
Evaluating the above integral for $d=3-\e$, we get
\be
16iN\lambda_{1}^{2}\frac{2\pi^{3}}{2\Gamma(2)(2\pi)^{4}}m^{2}_{1}(-p^{0}+\frac{\vec p^{2}}{4m_{1}})(-\frac{2}{\pi\e})=-\frac{2iN\lambda^{2}_{1}m^{2}_{1}}{\pi^{2}\e}(-p^{0}+\frac{\vec p^{2}}{4m_{1}})\,.
\ee
The contribution due to the counter term in the 2-point function is
\be
i\delta_{\sigma}(p^{0}-\frac{\vec p^{2}}{2m_{2}})\,.
\ee 
Cancelling the divergence, we obtain
\be
\delta_{\sigma}=-\frac{2N\lambda^{2}_{1}m^{2}_{1}}{\pi^{2}\e}\,.
\ee
Next, we calculate the 3-point function i.e. $<\sigma\phi^{I\dagger}\phi^{J\dagger}>$. However, at one loop we do not have any non trivial diagram. Next, we calculate the 4-point function. The relevant diagrams are shown in figure \ref{4-pointFn}\,.
The $\sigma$-loop gives the following contributions
\be
16\frac{1}{2}(i\lambda_{2})^{2}\int\frac{dE\,d^{d-1}k}{(2\pi)^{d}}G_{\sigma\sigma}(k+\frac{p_{s}}{2})G_{\sigma\sigma}(-k+\frac{p_{s}}{2})\,.
\ee
Here
\be
G_{\sigma\sigma}(k)=\frac{i}{k^{0}-\frac{\vec k^{2}}{4m_{1}}+i\delta}\,.
\ee
Evaluating the above in $d=3-\e$, we obtain
\be
4i\lambda_{2}^{2}\frac{2\pi}{(2\pi)^{2}}(2m_{1})\frac{\pi}{\cos\frac{\pi}{2}(1-\e)}=4i\lambda_{2}^{2}\frac{2\pi}{(2\pi)^{2}}(2m_{1})\frac{2}{\e}=\frac{8i\lambda^{2}_{2}m_{1}}{\pi\e}
\ee
There is one more contributions to 4-point function. It is due to scalar $\phi^{I}$ in the loop. The contribution is
\be
64N\frac{1}{4!}(i\lambda_{1})^{4}\int \frac{dEd^{d-1}k}{(2\pi)^{d}}G_{\p\p}(k+\frac{p_{1}}{2})G_{\p\p}(-k+\frac{p_{1}}{2})G_{\p\p}(p_{4}-k-\frac{p_{1}}{2})G_{\p\p}(p_{3}+k-\frac{p_{1}}{2})\,.
\ee
Evaluating the above contributions explicitly, we find that the integrals are convergent for $d=3-\e$.

Introducing the counter terms,
\be
m_{1}\lambda_{1}\sqrt{Z_{\sigma}}=\m^{\frac{\e}{2}}g_{1},\quad \lambda_{2}Z_{\sigma}^{2}m_{1}=\m^{\e}Z_{g_{2}}g_{2}\,,
\ee
 we have the contribution
\be
\frac{4i\m^{\e}}{m_{1}}g_{2}\delta_{g_{2}},\qquad\text{where}\quad Z_{g_{2}}=1+\delta_{g_{2}}\,.
\ee
The divergence is cancelled if we choose 
\be
\delta_{g_{2}}=-\frac{2g_{2}}{\pi\e}\,.
\ee
In terms of the renormalized coupling, the wave function renormalization is
\be
\delta_{\sigma}=-\frac{2Ng^{2}_{1}}{\pi^{2}\e}\,.
\ee
We have two $\beta$-function equation. 
\bea
\beta_{1}=-\frac{\e}{2}g_{1}+\frac{Ng^{3}_{1}}{\pi^{2}},\quad\text{and}\quad \beta_{2}=-\e g_{2}-\frac{2g_{2}^{2}}{\pi}+\frac{4g_{1}^{2}g_{2}N}{\pi^{2}}\,.
\eea
The locations of the IR interacting fixed points are
\bea
(g^{2}_{1},g_{2})=(\frac{\pi^{2}\e}{2N},0), (\frac{\pi^{2}\e}{2N},\frac{\pi\e}{2})\,.
\eea
\bibliographystyle{JHEP}

\begin{thebibliography}{10}

\bibitem{Hagen:1972pd}
C.~R. Hagen, {\it {Scale and conformal transformations in galilean-covariant
  field theory}},  {\em Phys. Rev. D} {\bf 5} (1972) 377--388.

\bibitem{Niederer:1972}
U.~Niederer, {\it {The maximal kinematical invariance group of the free
  Schroedinger equation}},  {\em Helvetica Physica Acta} {\bf 45} (1972) 802.

\bibitem{Henkel:1993sg}
M.~Henkel, {\it {Schrodinger invariance in strongly anisotropic critical
  systems}},  {\em J. Statist. Phys.} {\bf 75} (1994) 1023--1061,
  [\href{http://arxiv.org/abs/hep-th/9310081}{{\tt hep-th/9310081}}].

\bibitem{Mehen:1999nd}
T.~Mehen, I.~W. Stewart, and M.~B. Wise, {\it {Conformal invariance for
  nonrelativistic field theory}},  {\em Phys. Lett. B} {\bf 474} (2000)
  145--152, [\href{http://arxiv.org/abs/hep-th/9910025}{{\tt
  hep-th/9910025}}].

\bibitem{Henkel:2003pu}
M.~Henkel and J.~Unterberger, {\it {Schrodinger invariance and space-time
  symmetries}},  {\em Nucl. Phys. B} {\bf 660} (2003) 407--435,
  [\href{http://arxiv.org/abs/hep-th/0302187}{{\tt hep-th/0302187}}].

\bibitem{Nishida:2006br}
Y.~Nishida and D.~T. Son, {\it {An Epsilon expansion for Fermi gas at infinite
  scattering length}},  {\em Phys. Rev. Lett.} {\bf 97} (2006) 050403,
  [\href{http://arxiv.org/abs/cond-mat/0604500}{{\tt cond-mat/0604500}}].

\bibitem{Nishida:2006eu}
Y.~Nishida and D.~T. Son, {\it {Fermi gas near unitarity around four and two
  spatial dimensions}},  {\em Phys. Rev. A} {\bf 75} (2007) 063617,
  [\href{http://arxiv.org/abs/cond-mat/0607835}{{\tt cond-mat/0607835}}].

\bibitem{Nishida:2007pj}
Y.~Nishida and D.~T. Son, {\it {Nonrelativistic conformal field theories}},
  {\em Phys. Rev. D} {\bf 76} (2007) 086004,
  [\href{http://arxiv.org/abs/0706.3746}{{\tt arXiv:0706.3746}}].

\bibitem{Braaten:2008uh}
E.~Braaten and L.~Platter, {\it {Exact Relations for a Strongly Interacting
  Fermi Gas from the Operator Product Expansion}},  {\em Phys. Rev. Lett.} {\bf
  100} (2008) 205301, [\href{http://arxiv.org/abs/0803.1125}{{\tt
  arXiv:0803.1125}}].

\bibitem{Golkar:2014mwa}
S.~Golkar and D.~T. Son, {\it {Operator Product Expansion and Conservation Laws
  in Non-Relativistic Conformal Field Theories}},  {\em JHEP} {\bf 12} (2014)
  063, [\href{http://arxiv.org/abs/1408.3629}{{\tt arXiv:1408.3629}}].

\bibitem{Goldberger:2014hca}
W.~D. Goldberger, Z.~U. Khandker, and S.~Prabhu, {\it {OPE convergence in
  non-relativistic conformal field theories}},  {\em JHEP} {\bf 12} (2015) 048,
  [\href{http://arxiv.org/abs/1412.8507}{{\tt arXiv:1412.8507}}].

\bibitem{Affleck:1991tk}
I.~Affleck and A.~W.~W. Ludwig, {\it {Universal noninteger 'ground state
  degeneracy' in critical quantum systems}},  {\em Phys. Rev. Lett.} {\bf 67}
  (1991) 161--164.

\bibitem{Cardy:1991tv}
J.~L. Cardy and D.~C. Lewellen, {\it {Bulk and boundary operators in conformal
  field theory}},  {\em Phys. Lett. B} {\bf 259} (1991) 274--278.

\bibitem{McAvity:1993ue}
D.~McAvity and H.~Osborn, {\it {Energy momentum tensor in conformal field
  theories near a boundary}},  {\em Nucl. Phys. B} {\bf 406} (1993) 655--680,
  [\href{http://arxiv.org/abs/hep-th/9302068}{{\tt hep-th/9302068}}].

\bibitem{McAvity:1995zd}
D.~McAvity and H.~Osborn, {\it {Conformal field theories near a boundary in
  general dimensions}},  {\em Nucl. Phys. B} {\bf 455} (1995) 522--576,
  [\href{http://arxiv.org/abs/cond-mat/9505127}{{\tt cond-mat/9505127}}].

\bibitem{Friedan:2003yc}
D.~Friedan and A.~Konechny, {\it {On the boundary entropy of one-dimensional
  quantum systems at low temperature}},  {\em Phys. Rev. Lett.} {\bf 93} (2004)
  030402, [\href{http://arxiv.org/abs/hep-th/0312197}{{\tt
  hep-th/0312197}}].

\bibitem{Cardy:2004hm}
J.~L. Cardy, {\it {Boundary conformal field theory}},
  \href{http://arxiv.org/abs/hep-th/0411189}{{\tt hep-th/0411189}}.

\bibitem{Fursaev:2015wpa}
D.~Fursaev, {\it {Conformal anomalies of CFT\textquoteright{}s with
  boundaries}},  {\em JHEP} {\bf 12} (2015) 112,
  [\href{http://arxiv.org/abs/1510.01427}{{\tt arXiv:1510.01427}}].

\bibitem{Solodukhin:2015eca}
S.~N. Solodukhin, {\it {Boundary terms of conformal anomaly}},  {\em Phys.
  Lett. B} {\bf 752} (2016) 131--134,
  [\href{http://arxiv.org/abs/1510.04566}{{\tt arXiv:1510.04566}}].

\bibitem{Herzog:2015ioa}
C.~P. Herzog, K.-W. Huang, and K.~Jensen, {\it {Universal Entanglement and
  Boundary Geometry in Conformal Field Theory}},  {\em JHEP} {\bf 01} (2016)
  162, [\href{http://arxiv.org/abs/1510.00021}{{\tt arXiv:1510.00021}}].

\bibitem{Jensen:2015swa}
K.~Jensen and A.~O'Bannon, {\it {Constraint on Defect and Boundary
  Renormalization Group Flows}},  {\em Phys. Rev. Lett.} {\bf 116} (2016),
  no.~9 091601, [\href{http://arxiv.org/abs/1509.02160}{{\tt
  arXiv:1509.02160}}].

\bibitem{Herzog:2017xha}
C.~P. Herzog and K.-W. Huang, {\it {Boundary Conformal Field Theory and a
  Boundary Central Charge}},  {\em JHEP} {\bf 10} (2017) 189,
  [\href{http://arxiv.org/abs/1707.06224}{{\tt arXiv:1707.06224}}].

\bibitem{Casini:2018nym}
H.~Casini, I.~Salazar~Landea, and G.~Torroba, {\it {Irreversibility in quantum
  field theories with boundaries}},  {\em JHEP} {\bf 04} (2019) 166,
  [\href{http://arxiv.org/abs/1812.08183}{{\tt arXiv:1812.08183}}].

\bibitem{Kobayashi:2018lil}
N.~Kobayashi, T.~Nishioka, Y.~Sato, and K.~Watanabe, {\it {Towards a
  $C$-theorem in defect CFT}},  {\em JHEP} {\bf 01} (2019) 039,
  [\href{http://arxiv.org/abs/1810.06995}{{\tt arXiv:1810.06995}}].

\bibitem{Wang:2021mdq}
Y.~Wang, {\it {Defect $a$-Theorem and $a$-Maximization}},
  \href{http://arxiv.org/abs/2101.12648}{{\tt arXiv:2101.12648}}.

\bibitem{Duval:1984cj}
C.~Duval, G.~Burdet, H.~P. Kunzle, and M.~Perrin, {\it {Bargmann Structures and
  Newton-cartan Theory}},  {\em Phys. Rev. D} {\bf 31} (1985) 1841--1853.

\bibitem{Duval:2009vt}
C.~Duval and P.~A. Horvathy, {\it {Non-relativistic conformal symmetries and
  Newton-Cartan structures}},  {\em J. Phys. A} {\bf 42} (2009) 465206,
  [\href{http://arxiv.org/abs/0904.0531}{{\tt arXiv:0904.0531}}].

\bibitem{Jensen:2014aia}
K.~Jensen, {\it {On the coupling of Galilean-invariant field theories to curved
  spacetime}},  {\em SciPost Phys.} {\bf 5} (2018), no.~1 011,
  [\href{http://arxiv.org/abs/1408.6855}{{\tt arXiv:1408.6855}}].

\bibitem{Son:2005rv}
D.~T. Son and M.~Wingate, {\it {General coordinate invariance and conformal
  invariance in nonrelativistic physics: Unitary Fermi gas}},  {\em Annals
  Phys.} {\bf 321} (2006) 197--224,
  [\href{http://arxiv.org/abs/cond-mat/0509786}{{\tt cond-mat/0509786}}].

\bibitem{Son:2013rqa}
D.~T. Son, {\it {Newton-Cartan Geometry and the Quantum Hall Effect}},
  \href{http://arxiv.org/abs/1306.0638}{{\tt arXiv:1306.0638}}.

\bibitem{Geracie:2014nka}
M.~Geracie, D.~T. Son, C.~Wu, and S.-F. Wu, {\it {Spacetime Symmetries of the
  Quantum Hall Effect}},  {\em Phys. Rev. D} {\bf 91} (2015) 045030,
  [\href{http://arxiv.org/abs/1407.1252}{{\tt arXiv:1407.1252}}].

\bibitem{Herzog:2017kkj}
C.~Herzog, K.-W. Huang, and K.~Jensen, {\it {Displacement Operators and
  Constraints on Boundary Central Charges}},  {\em Phys. Rev. Lett.} {\bf 120}
  (2018), no.~2 021601, [\href{http://arxiv.org/abs/1709.07431}{{\tt
  arXiv:1709.07431}}].

\bibitem{Jensen:2014hqa}
K.~Jensen, {\it {Anomalies for Galilean fields}},  {\em SciPost Phys.} {\bf 5}
  (2018), no.~1 005, [\href{http://arxiv.org/abs/1412.7750}{{\tt
  arXiv:1412.7750}}].

\bibitem{Arav:2016xjc}
I.~Arav, S.~Chapman, and Y.~Oz, {\it {Non-Relativistic Scale Anomalies}},  {\em
  JHEP} {\bf 06} (2016) 158, [\href{http://arxiv.org/abs/1601.06795}{{\tt
  arXiv:1601.06795}}].

\bibitem{Auzzi:2015fgg}
R.~Auzzi, S.~Baiguera, and G.~Nardelli, {\it {On Newton-Cartan trace
  anomalies}},  {\em JHEP} {\bf 02} (2016) 003,
  [\href{https://arxiv.org/abs/1511.08150}{{\tt arXiv:1511.08150}}]. [Erratum:
  JHEP 02, 177 (2016)].

\bibitem{Auzzi:2017wwc}
R.~Auzzi, S.~Baiguera, and G.~Nardelli, {\it {Nonrelativistic trace and
  diffeomorphism anomalies in particle number background}},  {\em Phys. Rev. D}
  {\bf 97} (2018), no.~8 085010, [\href{http://arxiv.org/abs/1711.00910}{{\tt
  arXiv:1711.00910}}].

\end{thebibliography}
\providecommand{\href}[2]{#2}\begingroup\raggedright\endgroup

\end{document}